\documentclass[preprint,aps]{revtex4}
\newcommand {\bc}{\begin{center}}
\newcommand {\ec}{\end{center}}
\newcommand {\bea}{\begin{eqnarray}}
\newcommand {\eea}{\end{eqnarray}}
\newcommand {\be}{\begin{equation}}
\newcommand {\ee}{\end{equation}}

\def\lsim{\mathrel{\rlap{\lower4pt\hbox{\hskip1pt$\sim$}}
    \raise1pt\hbox{$<$}}}               
\def\gsim{\mathrel{\rlap{\lower4pt\hbox{\hskip1pt$\sim$}}
    \raise1pt\hbox{$>$}}}                
\usepackage{graphicx}

\begin{document}

%\draft

\title{Elliptic flow of the dilute Fermi gas: From kinetics to
hydrodynamics}

\author{K.~Dusling and T.~Sch\"afer}

\affiliation{Department of Physics, North Carolina State University,
Raleigh, NC 27695}

\begin{abstract}
We use the Boltzmann equation in the relaxation time approximation
to study the expansion of a dilute Fermi gas at unitarity. We focus, 
in particular, on the approach to the hydrodynamic limit. Our main 
finding are: i) In the regime that has been studied experimentally 
hydrodynamic effects beyond the Navier-Stokes approximation are small, 
ii) mean field corrections to the Boltzmann equation are not important, 
iii) experimental data imply that freezeout occurs very late, that means
that the relaxation time remains smaller than the expansion time 
during the entire evolution of the system, iv) the experimental
results also imply that the bulk viscosity is significantly smaller 
than the shear viscosity of the system. 

\end{abstract}

\maketitle

%%%%%%%%%%%%%%%%%%%%%%%%%%%%%%%%%%%%%%%%%%%%%%%%%%%%%%%%%%%%%%%%%%%%%%%%
\section{Introduction}
\label{sec_intro}
%%%%%%%%%%%%%%%%%%%%%%%%%%%%%%%%%%%%%%%%%%%%%%%%%%%%%%%%%%%%%%%%%%%%%%%%

 The dilute Fermi gas at unitarity is a strongly correlated 
quantum fluid that serves as a new paradigm for strong correlations
in many other systems, like dilute neutron matter, the quark-gluon 
plasma, and high $T_c$ superconductors
\cite{Bloch:2007,Giorgini:2008,Chin:2009,Schafer:2009dj}.
The interest in the unitary Fermi gas derives in large part 
from the fact that the system provides a particularly simple 
realization of strong correlations. At unitarity the $s$-wave
scattering length is infinite, and the effective range and
all other scattering parameters are zero. This implies that 
even though the system is very strongly correlated, details 
of the interaction are not important and the fluid is scale 
invariant. 

 An important manifestation of strong correlations is the observation 
of nearly ideal hydrodynamic flow \cite{OHara:2002}. Recently, a
significant amount of effort has been devoted to quantifying this 
observation by determining the shear viscosity $\eta$ or the 
dimensionless ratio of shear viscosity to entropy density $\eta/s$ 
\cite{Schafer:2007pr,Turlapov:2007,Schaefer:2009px,Thomas:2009zz,Cao:2010}.
This effort is inspired, in part, by the conjecture that there 
is a universal lower bound on the shear viscosity to entropy 
density of strongly coupled fluids \cite{Kovtun:2004de},
and by similar measurements of $\eta/s$ for the quark gluon plasma
created in heavy ion collisions at RHIC and the LHC 
\cite{rhic:2005,Romatschke:2007mq,Dusling:2007gi,Song:2008hj}.

 Measurements of the shear viscosity of the unitary Fermi gas are 
based on using viscous hydrodynamics to analyze the expansion of 
a gas cloud after the confining potential is removed. Alternatively, 
one can study collective oscillations of the cloud. The main difficulty 
with these analyses is that hydrodynamics breaks down in the dilute 
corona of the trapped gas, and that dissipative corrections from this 
regime can potentially be large. It is therefore important to study 
the crossover from weakly collisional kinetic dynamics to almost ideal 
hydrodynamics. 

 In this paper we use the Boltzmann (and Boltzmann-Vlasov)
equation in the relaxation time approximation to study a number
of issues related to the crossover from kinetics to hydrodynamics. 
In Sect.~\ref{sec_relax} we discuss the matching between the 
Boltzmann equation and the Navier-Stokes equation for a simple 
functional form of the relaxation time. We also study, for the 
system parameters realized in the recent experiments of Cao 
et al.~\cite{Cao:2010}, the magnitude of hydrodynamic effects 
beyond the Navier-Stokes approximation. In Sect.~\ref{sec_mfa}
we investigate the role of mean field effects in the 
Boltzmann-Vlasov equation. In Sect.~\ref{sec_freeze} we study 
the question whether there is any experimental evidence for 
freezeout, that is a transition between hydrodynamic and ballistic
expansion during the evolution of the system. In 
Sect.~\ref{sec_kin_rel} we study a more realistic model of the 
relaxation time, and in Sect.~\ref{sec_bulk} we study the 
possible role of bulk viscosity. We should note that there is 
an extensive literature on using the Boltzmann equation to
describe the dynamics of trapped Bose and Fermi gases, see for 
example \cite{Guery:1999,Pedri:2002,Guery:2002,Menotti:2002,Jackson:2004,Nikuni:2004,Bruun:2005,Bruun:2006,Bruun:2007}. In this paper we focus, in 
particular, on the questions how close recent experiments are 
to the hydrodynamic limit, and how large the associated uncertainties
in the shear viscosity are.

%%%%%%%%%%%%%%%%%%%%%%%%%%%%%%%%%%%%%%%%%%%%%%%%%%%%%%%%%%%%%%%%%%%%%%%%
\section{Boltzmann equation in relaxation time approximation}
\label{sec_relax}
%%%%%%%%%%%%%%%%%%%%%%%%%%%%%%%%%%%%%%%%%%%%%%%%%%%%%%%%%%%%%%%%%%%%%%%%

 We consider the limit in which the unitary Fermi gas can be
described in terms of a quasi-particle distribution function 
$f(x,v,t)$. The equation of motion for the distribution 
function is the Boltzmann-Vlasov equation
\be
\label{boltz} 
 \frac{\partial f}{\partial t} 
   + v_i \frac{\partial f}{\partial x_i} 
   - \frac{1}{m} \frac{\partial (U_{\it mf}+U_{\it ext})}{\partial x_i} 
      \frac{\partial f}{\partial v_i} = C[f]\, , 
\ee
where $U_{\it ext}=(m/2)\omega_i^2x_i^2$ is the external 
confinement potential, $U_{\it mf}$ is a mean field potential, 
and $C[f]$ is the collision term. We will study the effect of the 
mean field potential in Sect.~\ref{sec_mfa}. For now we will 
set  $U_{\it mf}=0$. Throughout this paper we will approximate the
collision term using the relaxation time approximation
\be 
\label{c_relax}
 C[f] \simeq - \frac{f-f_{\it le}}{\tau}\, , 
\ee
where $\tau$ is the relaxation time and $f_{\it le}$ is the 
local equilibrium distribution. The relaxation time does not
have to be a constant, and we will study different scaling 
laws for $\tau$ below.

 We are interested in the time evolution of a harmonically 
trapped cloud after the trapping potential is removed. We 
will follow \cite{Pedri:2002,Guery:2002} and use a 
scaling ansatz for the distribution function
\be 
\label{f_scal_ans}
 f(x,v,t) = \Gamma(t) f_0(R(t),U(t))
\ee
with 
\be 
\label{R_i_def}
R_i= \frac{x_i}{b_i}\, ,\hspace{0.5cm}
U_i= \frac{v_i-\frac{\dot{b}_i}{b_i}x_i}{\theta_i^{1/2}}\, ,\hspace{0.5cm}
\Gamma = \prod_j\frac{1}{b_j\theta_j^{1/2}}\, ,
\ee
where $b_i,\theta_i$ are functions of $t$ and $f_0(r,v)$ is a 
solution of the Boltzmann-Vlasov equation in equilibrium
\be
\label{boltz_0} 
     v_i \frac{\partial f_0}{\partial x_i} 
   = \frac{1}{m} \frac{\partial (U_{\it mf}+U_{\it ext})}{\partial x_i}
      \frac{\partial f_0}{\partial v_i} \, . 
\ee
In the absence of a mean field potential this equation is solved
by a Maxwell distribution. The scaling ansatz (\ref{f_scal_ans}) 
breaks local thermal equilibrium only through the anisotropy of 
the temperature parameters $\theta_i$. The corresponding local 
equilibrium distribution $f_{\it le}$ can be found by replacing 
$\theta_i\to \bar{\theta}=(\sum_i\theta_i)/3$. This distribution
function is characterized by having the same mean kinetic energy
as the non-equilibrium distribution $f$. In the free streaming 
limit $\tau\to\infty$ equ.~(\ref{f_scal_ans}) solves the Boltzmann
equation exactly. In the presence of collisions we can obtain a
differential equation for the parameters $b_i(t)$ and $\theta_i(t)$ 
by computing moments of the Boltzmann equation. Integrating the 
Boltzmann equation over $\int d^3U\,d^3R\, U_jR_j$ (no sum 
over $j$) gives \cite{Pedri:2002}
\be 
\label{BE_b_j}
\ddot{b}_j + \omega_j^2 b_j 
   - \omega_j^2 \frac{\theta_j}{b_j} = 0 \, . 
\ee
The second term is due to the external potential and is not
present if one considers an expanding cloud. Taking moments
of the form  $\int d^3U\,d^3R\, U_jU_j$ gives
\be 
\label{BE_th_j}
\dot{\theta}_j + 2\frac{\dot{b}_j}{b_j}\theta_j = 
 - \frac{1}{\tau} \left(\theta_j-\bar\theta\right) \, . 
\ee
Taking moments weighted with $R_jR_j$ does not provide additional 
constraints. Together with the initial conditions $b_j(0)=\theta_j(0)=1$ 
and $\dot{b}_j(0)=0$ equations (\ref{BE_b_j},\ref{BE_th_j}) describe 
the evolution of an expanding cloud.

 We expect the Boltzmann equation to reduce to hydrodynamics in 
the limit that the relaxation time is shorter than the expansion
time, $\tau\ll \tau_{\it exp} = (\sum_i \dot{b}_i/b_i)^{-1}$. 
In the limit $\tau\to 0$ equ.~(\ref{BE_th_j}) implies that 
$\theta_i=\bar{\theta}$ and $\bar{\theta}=(\prod_i b_i)^{-2/3}$. 
The equation of motion for the scale parameter of an 
expanding cloud reduces to
\be 
\label{b_i_euler}
\ddot{b}_i - \frac{\omega_i^2}{b_i\prod_j b_j^{2/3}} = 0\, .
\ee
This equation is identical to the equation of motion that 
follows from the Euler equation in ideal hydrodynamics 
\cite{Menotti:2002,Schaefer:2009px}. If we keep terms of order 
$\tau$ we recover dissipative terms in hydrodynamics. For this 
purpose we compute the conserved charges (mass, momentum, and 
energy) and the associated currents. The mass density is given 
by $\rho=mn=m\int d^3v f(x,v,t)$. The  momentum density is 
\be
\label{g_i}
g_i = m\int d^3v\, v_i f(x,v,t) = \rho V_i \,  
\ee
where $V_i=(\dot{b}_i/b_i)x_i$ is the local fluid velocity. 
The energy density is 
\be 
{\cal E} = \frac{1}{2} \int d^3v\, mv^2f(x,v,t) 
   = \frac{1}{2}\rho V^2 + n\epsilon_0\, , 
\ee
with $\epsilon_0 =\frac{3}{2}T_{\it le}$. The stress tensor is 
given by 
\be  
\label{pi_ij_f}
\Pi_{ij} = \int d^3v\, m  v_iv_j f(x,v,t) = 
    \rho V_iV_j + \delta_{ij}\rho \theta_i \langle v_i^2\rangle \, ,  
\ee
where $\langle v_i^2\rangle = {\cal V}^{-1} \int d^3v \, f_0(R,v)$ with 
${\cal V}=\prod_i b_j$ is an average with respect to the equilibrium 
distribution in the local rest frame. We can write $f=f_{\it le}+\delta f$ 
and split equ.~(\ref{pi_ij_f}) into a local equilibrium and 
dissipative part. The dissipative part is 
\be 
\label{del_pi_1}
\delta\Pi_{ij} =-\delta_{ij} n T_{\it le} 
  \left( 1 - \frac{\theta_i}{\bar\theta} \right)\, . 
\ee
We can use equ.~(\ref{BE_th_j}) to express $\theta_i/\bar\theta$ 
in terms of $\theta_i$ and $b_i$. The result is explicitly 
proportional to $\tau$, so that at leading order we can 
use the relation between $\theta_i$ and $b_i$ in ideal 
hydrodynamics. We find
\be 
\label{del_pi_2}
\delta \Pi_{ij} = -\delta_{ij} nT_{\it le} \tau 
  \left( 2\frac{\dot{b_i}}{b_i} 
        - \frac{2}{3}\sum_k  \frac{\dot{b_k}}{b_k} \right)\, .
\ee
Using $V_i=x_i(\dot{b}_i/b_i)$ we see that equ.~(\ref{del_pi_2}) matches 
the viscous stress tensor in hydrodynamics, $\delta \Pi_{ij}=-\eta
(\partial_iV_j+\partial_j V_i-\frac{2}{3}\delta_{ij}(\partial\cdot V))
-\zeta\delta_{ij}(\partial\cdot V)$, if we set 
\be 
\label{eta_relax}
 \eta = n\tau T_{\it le}\, , 
\ee
and $\zeta=0$. Equ.~(\ref{eta_relax}) is a standard result in kinetic 
theory \cite{Bruun:2005}. The bulk viscosity is expected to vanish 
in the dilute Fermi gas at unitarity \cite{Son:2005tj,Enss:2010qh},
but we will see that the relaxation time approximation leads to 
a vanishing bulk viscosity even in cases where we expect the 
physical result to be non-zero. Scale invariance also implies that 
$\eta=n \alpha_n(y)$, where $\alpha_n$ is a function of $y=mT/n^{2/3}$, 
and $\tau=\tau_n(y)/T$. The matching condition (\ref{eta_relax}) then 
implies that $\tau_n(y)=\alpha_n(y)$. The simplest case is that $\alpha_n$ 
is a constant. In this case the relaxation time is only a function 
of temperature $\tau=\alpha_n/T_{\it le}=\alpha_n/(T_0\bar\theta)$, 
where $T_0$ is the initial temperature.

%%%%%%%%%%%%%%%%%%%%%%%%%%%%%%%%%%%%%%%%%%%%%%%%%%%%%%%%%%%%%%%%%%%%%%%%%
\begin{figure}[t]
\bc\includegraphics[width=10cm]{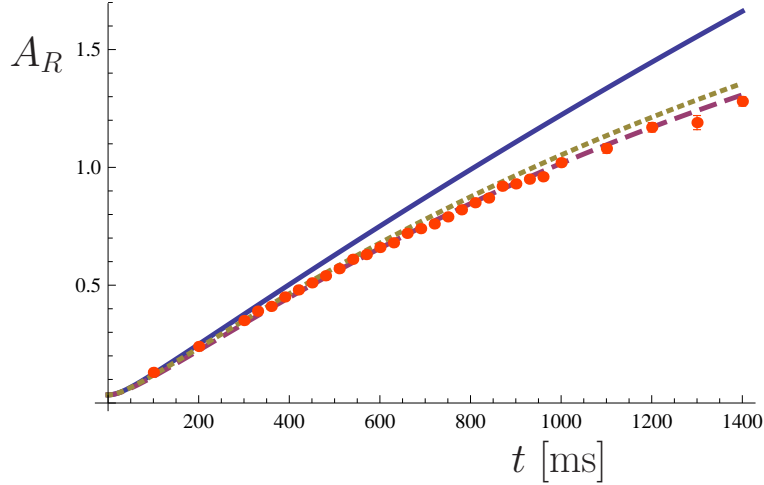}\ec
\caption{\label{fig_relax_vs_hydro}
This figure shows the matching between kinetic theory and 
Navier-Stokes hydrodynamics. We show the evolution of the 
aspect ratio $A_R$ as a function of time. The solid points 
are data taken at an initial energy $E/E_F=3.61$ \cite{Cao:2010}.
The solid line shows a solution of the Euler equation, the 
long dashed line is a solution of the Navier-Stokes equation
where the viscosity coefficient $\alpha_n=22.1$ ($\eta=\alpha_n n$)
was adjusted to reproduce the data, and the short-dashed
line is a solution to the Boltzmann equation in the relaxation
time approximation with $\tau=\alpha_n/T$.}   
\end{figure}
%%%%%%%%%%%%%%%%%%%%%%%%%%%%%%%%%%%%%%%%%%%%%%%%%%%%%%%%%%%%%%%%%%%%%%%%%

 In Fig.~\ref{fig_relax_vs_hydro} we compare the solution of the 
Boltzmann equation to a solution of the Navier-Stokes equation.
The Navier-Stokes solution is described in \cite{Schafer:2010dv},
and the parameter $\alpha_n$ was fitted to data taken by Cao
et al.~\cite{Cao:2010} at an initial energy $E/E_F=3.61$, where 
$E_F=(3N)^{1/3}\bar\omega$. The experiment involves an axially 
symmetric trap with $\omega_\perp\equiv\omega_x=\omega_y$. The trap
frequencies are $\omega_\perp/(2\pi)=5078.7$ Hz and $\omega_z/(2\pi)
=175.7$ Hz, the number of particles is $N=2.61\cdot 10^5$ and we 
have defined $\bar\omega=\lambda^{1/3}\omega_\perp$ with $\lambda=
\omega_z/\omega_\perp$. The relaxation time was fixed according to 
the relation $\tau_n=\alpha_n$. Fig.~\ref{fig_relax_vs_hydro} shows 
the time evolution of the aspect ratio $A_R(t)=b_\perp(t)/b_z(t)$. 
We observe that the difference between the kinetic theory calculation 
and Navier-Stokes hydrodynamics is indeed small, indicating that in 
the regime that has been studied experimentally higher order hydrodynamic 
effects are small. 

%%%%%%%%%%%%%%%%%%%%%%%%%%%%%%%%%%%%%%%%%%%%%%%%%%%%%%%%%%%%%%%%%%%%%%%%%
\begin{figure}[t]
\bc\includegraphics[width=10cm]{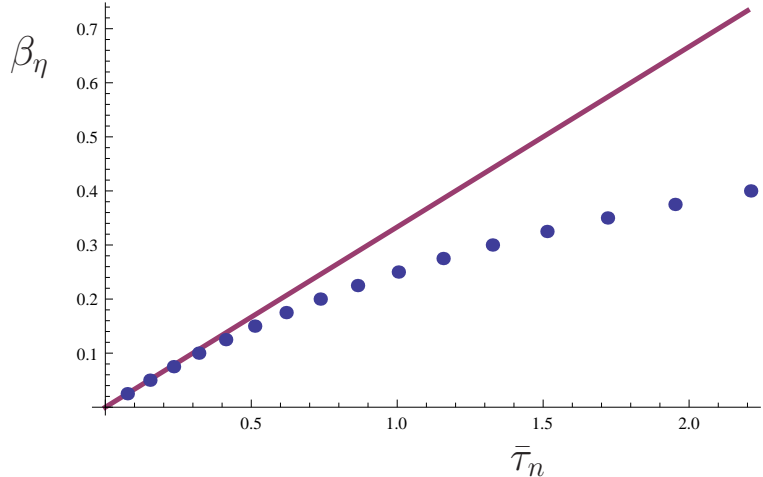}\ec
\caption{\label{fig_eta_vs_tau}
This figure shows the relation between the relaxation time in kinetic
theory and the shear viscosity. We plot the dimensionless relaxation
time $\bar\tau$ against the dimensionless shear viscosity $\beta_\eta$
defined in the text. The points were obtained by first solving the 
Navier-Stokes equation with a given values of $\beta_\eta$ for a 
trapped cloud with initial aspect ratio $A_R(0)=0.035$, and then 
determining the value of $\bar\tau$ that leads to the best fit 
of the solution of the Boltzmann equation to the Navier-Stokes
result for $A_R(t)$ in the regime $t<1400$ ms. The line show the 
leading order relation $\beta_\eta=\bar{\tau}/3$. Note that the 
long dashed line in Fig.~\ref{fig_relax_vs_hydro} corresponds 
to $\beta_\eta=0.25$.}   
\end{figure}
%%%%%%%%%%%%%%%%%%%%%%%%%%%%%%%%%%%%%%%%%%%%%%%%%%%%%%%%%%%%%%%%%%%%%%%%%

 This issue is studied in more detail in Fig.~\ref{fig_eta_vs_tau}. 
For a given value of the viscosity coefficient $\alpha_n$ we 
determine the Navier-Stokes evolution and then fit the relaxation 
time coefficient $\tau_n$ in the kinetic calculation to provide
the best fit to hydrodynamics. The figure shows the dimensionless
parameter 
\be 
\label{beta_eta}
 \beta_\eta = \frac{\alpha_n}{(3N\lambda)^{1/3}}\frac{1}{E_0/E_F}\, 
\ee
that governs the hydrodynamic evolution as a function of the dimensionless
relaxation time 
\be 
\bar{\tau}_n= \frac{\omega_\perp\tau_n}{T_0}\, . 
\ee 
In the hydrodynamic limit equ.~(\ref{eta_relax}) implies $\beta_\eta=
\bar\tau_n/3$. This equation is satisfied at small $\tau_n$, but for 
$\bar\tau_n\gsim 1$ the effective shear viscosity is smaller than the
one predicted by the linear approximation. The fit in 
Fig.~\ref{fig_relax_vs_hydro} corresponds to $\bar\tau_n=0.78$, which 
is close to the regime where higher order effects become large.  
Qualitatively, the behavior of $\beta_\eta$ as a function of $\bar\tau_n$
is easy to understand. At moderate $\tau_n$ the relaxation time approximation
implies that the dissipative stress tensor equ.~(\ref{del_pi_1}) does 
not reach the hydrodynamic limit equ.~(\ref{del_pi_2}) instantaneously,
but over a characteristic time $\tau_n/T_{\it le}$. For an expanding 
system this means that the time integral of the dissipative stresses
in the kinetic theory is smaller as compared to the hydrodynamic limit. 
At large $\tau_n$ the solution of the Boltzmann equation approaches 
the ballistic limit, and dissipative effects saturate as $\bar\tau_n
\to\infty$. 

%%%%%%%%%%%%%%%%%%%%%%%%%%%%%%%%%%%%%%%%%%%%%%%%%%%%%%%%%%%%%%%%%%%%%%%%
\section{Mean field effects}
\label{sec_mfa}
%%%%%%%%%%%%%%%%%%%%%%%%%%%%%%%%%%%%%%%%%%%%%%%%%%%%%%%%%%%%%%%%%%%%%%%%

%%%%%%%%%%%%%%%%%%%%%%%%%%%%%%%%%%%%%%%%%%%%%%%%%%%%%%%%%%%%%%%%%%%%%%%%%
\begin{figure}[t]
\bc\includegraphics[width=10cm]{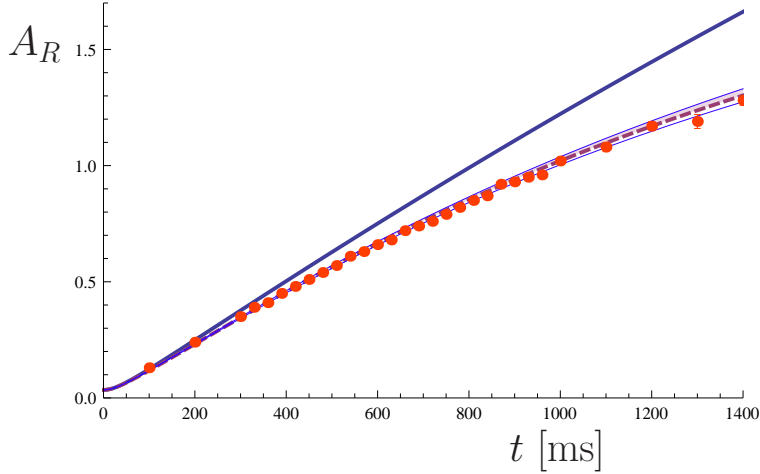}\ec
\caption{\label{fig_mf}
This figure shows the effect of the mean field potential on the 
evolution of the aspect ratio $A_R(t)$. The solid line shows the 
result in ideal hydrodynamics, and the points are experimental 
data from \cite{Cao:2010} taken at an initial energy $E/E_F=3.61$. 
The short dashed line is the result of a relaxation time fit to 
the data without mean field effects. The band around the dashed
line shows the variation of $A_R(t)$ when mean field effects 
included. The upper and lower limits of the band correspond to 
$\xi=\pm 0.1$, respectively.}   
\end{figure}
%%%%%%%%%%%%%%%%%%%%%%%%%%%%%%%%%%%%%%%%%%%%%%%%%%%%%%%%%%%%%%%%%%%%%%%%%

 In this section we investigate possible mean field effects in the 
Boltzmann equation, in particular the question to what extent these
effects cause uncertainties in the extraction of the relaxation time.
Without mean field effects the equation of state is $P=nT$ and the 
relation between energy density and pressure is ${\cal E}=\frac{3}{2}
P$. In the unitary gas scale invariance restricts the equation of 
state to be of the form $P=nTh(y)$ with $y=mT/n^{2/3}$ and $h(y)$ is 
a universal function which approaches unity in the high temperature, 
low density limit. The relation ${\cal E}=\frac{3}{2}P$ is exact
irrespective of the functional form of $h(y)$. Scale invariance 
restricts the form of the mean field potential to be  $U_{mf}=g(y)
n(x)^{\alpha}$ with $\alpha=2/3$. For simplicity we will assume 
that $g(y)\sim{\it const}$. The mean field potential does not 
modify the equation for $\theta_j$. The equation for $b_j$ becomes 
\be 
\label{BV_b_j}
\ddot{b}_j + \omega_j^2 b_j 
   - \omega_j^2 \frac{\theta_j}{b_j}
   + \omega_j^2 \xi \left(\frac{\theta_j}{b_j}
       - \frac{1}{b_j{\cal V}^\alpha} \right) = 0 \, . 
\ee
with ${\cal V}=\prod_j b_j$. The parameter $\xi$ is defined by
\be
\label{xi_def} 
\xi =\frac{\langle U_{mf}\rangle }
  {\frac{1+\alpha}{3\alpha}m\langle v^2\rangle + \langle U_{mf}\rangle }\, .
\ee
In the limit of ideal hydrodynamics ($\tau\to 0$) we have $\theta_j
=\bar\theta={\cal V}^{-2/3}$. This implies that mean field effects 
do not modify the evolution a scale invariant, non-dissipative gas.
This result is known from studies of the unitary gas in the context 
of ideal hydrodynamics \cite{Schaefer:2009px,Schafer:2010dv}. The 
basic observation is that the Euler equation only depends on the 
relation $P({\cal E})$, and not on the equation of state $P(n,T)$.

For a finite relaxation time the relation $\theta_j={\cal V}^{-2/3}$ 
is only approximately satisfied and mean field corrections play 
a role. In order to estimate the size of mean field effects we 
relate the parameter $\xi$ to thermodynamic properties of the 
unitary gas. The static Boltzmann-Vlasov equation (\ref{boltz_0})
implies
\be
\label{Virial}
\frac{1}{2}m\langle v^2\rangle  
 + \frac{3}{2}\frac{\alpha}{1+\alpha} \langle U_{mf}\rangle 
  = \frac{m}{2} \sum_i \omega_i^2 \langle r_i^2 \rangle \, . 
\ee
For a scale invariant gas with $\alpha=2/3$ the LHS of this relation 
is equal to the internal energy of the gas. Equ.~(\ref{Virial}) is then 
the well-known Virial theorem \cite{Thomas:2005,Werner:2006,Son:2007}: 
The total internal energy of a harmonically trapped system is equal to 
the potential energy. We can use equ.~(\ref{Virial}) to express $\xi$ in 
terms of the total energy of the gas. For $\alpha=2/3$ and in the 
limit $\xi\ll 1$ we find
\be 
\label{xi_eos}
\xi =  \frac{E_0}{3NT} - 1 \, . 
\ee
For an ideal gas the total (internal plus potential) energy is 
$E_0=3NT$ and $\xi=0$. We have computed $\xi$ using the equation of 
state described in the appendix of \cite{Schafer:2010dv}. This
equation of state is a parameterization of the recent experimental
results of Nascimbene et al.~\cite{Chevy:2009}. We find that in 
the normal fluid regime $|\xi|<0.1$, where $\xi<0$ at high 
temperature regime and $\xi>0$ near the critical temperature 
$T_c$. We note that at unitarity corrections to the equation of 
state due to quantum statistics are parametrically of the same 
order as mean field corrections. At high temperature quantum
corrections are numerically small, but near $T_c$ equ.~(\ref{xi_eos})
is not reliable. 

 In Fig.~\ref{fig_mf} we show the effect of mean field corrections
in the regime $-0.1\leq\xi\leq 0.1$ on the evolution of the system.
Positive values of $\xi$ tend to push the evolution closer to the 
solution in ideal hydrodynamics, and therefore act like an effective
negative viscosity. Negative values of $\xi$ act an effective positive 
shear viscosity. Fig.~\ref{fig_mf} shows that mean field effects are 
small compared to dissipative effects. We can quantify this observation 
using the data at $E/E_F=3.61$. A fit of the relaxation time using the
Boltzmann equation without mean field effects gives $\bar\tau_n=1.14$. 
For this energy the equation of state described in \cite{Schafer:2010dv} 
gives $\xi=-0.014$. If we refit the relaxation time using the 
Boltzmann-Vlasov equation with this value of $\xi$ we find $\bar\tau_n
=1.12$, which is a $\sim 2\%$ correction.

%%%%%%%%%%%%%%%%%%%%%%%%%%%%%%%%%%%%%%%%%%%%%%%%%%%%%%%%%%%%%%%%%%%%%%%%
\section{Freezeout}
\label{sec_freeze}
%%%%%%%%%%%%%%%%%%%%%%%%%%%%%%%%%%%%%%%%%%%%%%%%%%%%%%%%%%%%%%%%%%%%%%%%

%%%%%%%%%%%%%%%%%%%%%%%%%%%%%%%%%%%%%%%%%%%%%%%%%%%%%%%%%%%%%%%%%%%%%%%%%
\begin{figure}[t]
\bc\includegraphics[width=10cm]{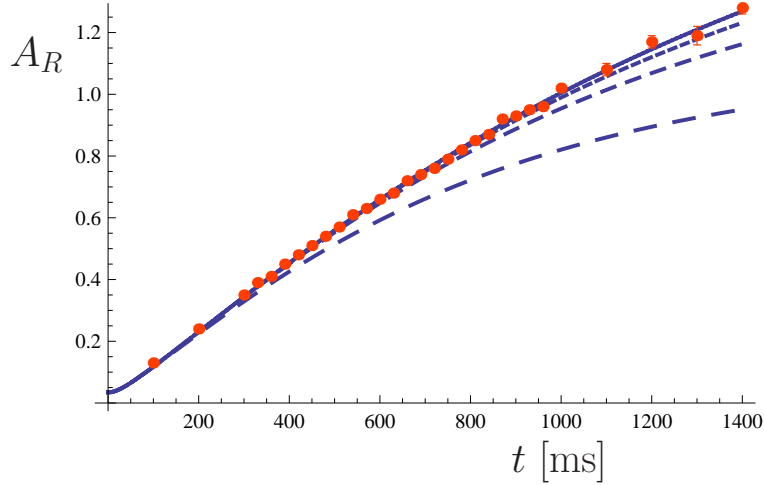}\ec
\caption{\label{fig_freeze}
This figure shows the effect of freezeout on the time evolution 
of the aspect ratio. The data points show the results of Cao et 
al.~also shown in Figs.~\ref{fig_relax_vs_hydro} and \ref{fig_mf}.
The solid line is the relaxation time fit shown in Fig.~\ref{fig_mf}. 
The dashed lines show the same fit but with the scaling of the 
relaxation time changed from $\tau\sim 1/T$ to $\tau\sim T^{1/2}
{\cal V}$ at a freezeout scale factor $b_{\it fr}=10,5,1$ (from 
top to bottom).}   
\end{figure}
%%%%%%%%%%%%%%%%%%%%%%%%%%%%%%%%%%%%%%%%%%%%%%%%%%%%%%%%%%%%%%%%%%%%%%%%%

 Up to this point we have considered a relaxation time of the form
$\tau=\tau_n/T$ with $\tau_n={\it const}$. In this case the relaxation
time scales as $\tau\sim \theta^{-1} \sim {\cal V}^{2/3}$. The 
expansion time, on the other hand, scales as $\tau\sim {\cal V}^{1/2}$
(at very late time the expansion changes from being two-dimensional
to three-dimensional and $\tau \sim {\cal V}^{1/3}$). This implies
that the system will eventually freeze out, and the nature of the 
expansion will change from hydrodynamic to ballistic. We note,
however, that for the values of $\tau_n$ that fit the data shown 
in Fig.~\ref{fig_relax_vs_hydro},\ref{fig_mf} freezeout occurs 
late, for $A_R>1$. At this time the density has dropped by a factor
of order $\lambda^{-2}\sim 10^3$.

 To maintain hydrodynamic behavior despite the large drop in density 
requires the system to be very close to the unitary limit. At unitarity
the mean free path scales as $l_{\it mfp}\sim (n\sigma)^{-1} \sim 
{\cal V}\lambda_{\it deB}^{-2} \sim {\cal V}^{1/3}$, where $\lambda_{\it deB}
\sim (mT)^{-1/2}$ is the thermal wave length. This implies that the 
the mean free path grows more slowly as the system size. For a 
weakly collisional system, on the other hand, we have $l_{\it mfp}\sim
(na^2)^{-1}\sim {\cal V}$ and the mean free path grows more quickly 
than the system size. 

 In this section we wish to check whether the data support the idea 
that freezeout occurs late. For this purpose we impose a freezeout 
time $t_{\it fr}$ so that for $t<t_{\it fr}$ the relaxation time scales 
as $\tau\sim \tau_n/T$ and for $t>t_{\it fr}$ the relaxation time scales 
as $\tau\sim \gamma {\cal V}/T^{1/2}$, which is the behavior expected 
for a constant cross section $\sigma\sim a^2$. The parameter $\tau_n$ 
is taken from the previous fit and $\gamma$ is adjusted so that $\tau$ 
is continuous at $t=t_{\it fr}$. In Fig.~\ref{fig_freeze} we show the 
evolution of the aspect ratio $A_R(t)$ for different choices of the 
freezeout time. The solid line shows the result for $t_{\it fr}\to\infty$, 
and the dashed curves correspond to freezeout scale parameters $b_\perp(
t_{\it fr})=10,5,1$. We observe that $t_{\it fr}\to\infty$ provides the best 
fit to the data, and $b_\perp(t_{\it fr})\lsim 5$ is clearly excluded. We 
conclude that for the conditions explored experimentally ($N\sim 2\cdot 
10^5$ and $E/E_F\leq 3.61$) the density can drop by a least a factor 
$\sim 25$ before freezeout occurs.

%%%%%%%%%%%%%%%%%%%%%%%%%%%%%%%%%%%%%%%%%%%%%%%%%%%%%%%%%%%%%%%%%%%%%%%%
\section{High temperature limit of the relaxation time}
\label{sec_kin_rel}
%%%%%%%%%%%%%%%%%%%%%%%%%%%%%%%%%%%%%%%%%%%%%%%%%%%%%%%%%%%%%%%%%%%%%%%%

  In the high temperature limit the viscous relaxation time can be 
computed reliably. The result is $\tau = \eta/(nT)$ with $\eta= 
\eta_0 (mT)^{3/2}$ and $\eta_0 = 15/(32\sqrt{\pi})$ 
\cite{Bruun:2005,Bruun:2006,Chao:2010tk}. This implies that 
even within the relaxation time approximation the collision
term is a nonlinear functional of the distribution function. 
We have 
\be 
\label{C_kin}
C[f] = - \frac{T_{\it le}}{\eta_0(mT_{\it le})^{3/2}} 
   \left[ \int d^3\bar{v}\, f(x,{\bar v},t) \right]
   \left( f(x,v,t) - f_{\it le}(x,v,t) \right)\, . 
\ee
We note that even in the simplest case $1/\tau\sim T_{\it le}/\tau_n$
with $\tau_n={\it const}$ the relaxation time is a functional
(via the temperature) of the distribution function. However, 
for the scaling ansatz in equ.~(\ref{f_scal_ans}) the temperature 
is only a function of time. The collision term in equ.~(\ref{C_kin})
is more complicated because the relaxation time is a function of 
time and position (but not of velocity). This means that it is 
difficult to find exact solutions of the Boltzmann equation. 
We obtain an approximate scaling solution by using the same 
ansatz for $f(x,v,t)$ as before. The parameters $b_i$ and 
$\theta_i$ are again fixed by taking moments of the Boltzmann
equation weighted with $R_i^2,U_i^2$ and $R_iU_i$. We find 
that only the $U_i^2$ equation is modified. As a consequence
equ.~(\ref{BE_th_j}) is replaced by
\be 
\label{BE_th_j_2}
\dot{\theta}_j + 2\frac{\dot{b}_j}{b_j}\theta_j = 
 - \frac{\langle n_0\rangle \bar\theta}{\eta_0 (m\bar\theta)^{3/2}{\cal V}} 
   \left(\theta_j-\bar\theta\right) \, , 
\ee
where $\langle n_0\rangle= \int d^3r\, n_0(r)^2/\int d^3r n_0(r)$ is the 
average density in the initial state. Note that for a Gaussian distribution
$\langle n_0\rangle = 2^{-3/2}n_0(0)$. We also observe that to leading order 
in $\tau$ the parameter $\bar\theta$ evolves as $\bar\theta={\cal V}^{-2/3}$. 
This implies that the RHS of equ.~(\ref{BE_th_j_2}) scales as $T_0\bar\theta/
\tau_n$ with $\tau_n=\eta_0 (mT_0)^{3/2}/\langle n_0\rangle$. As a 
consequence, the evolution of the system in the Navier-Stokes regime 
(that is to first order in $\tau$) is indistinguishable from the 
result shown in Fig.~\ref{fig_relax_vs_hydro}. What is new is the 
explicit relation between $\tau_n$ and the shear viscosity in 
the high temperature, low density limit. 

 This relation can be compared to the analysis presented by Cao et
al.~in \cite{Cao:2010}. In this work the authors attempted to extract 
$\eta_0$ from experimental data on elliptic flow in the high temperature 
regime using the Navier-Stokes equation. The problem with the Navier-Stokes 
analysis is that a density independent shear viscosity leads to an infinitely 
large dissipative contribution from the dilute corona. Cao et al.~argued that 
a correct treatment of the corona has to take into account the fact that at 
low density the dissipative stress tensor cannot instantaneously relax to 
the Navier-Stokes form. They proposed that this effect can be taken 
into account by using a simple model for the viscosity which is of the 
form $\eta(x)=\eta_0 (mT)^{3/2}(n(x)/n(0))$. A more detailed justification
for this model can be found in \cite{Schaefer:2009px}. The model of 
Cao et al.~can be written as $\eta=\alpha_n n$ with $\alpha_n=\eta_0
(mT)^{3/2}/n(0)$. Because the temperature drops (approximately) as 
$T\sim n(0)^{2/3}$ this corresponds to a constant value of $\alpha_n
=\eta_0 (mT_0)^{3/2}/n_0(0)$. This equation for $\alpha_n$ has the same 
structure as the result that follows from equ.~(\ref{BE_th_j_2}), but 
it differs by the numerical factor $\langle n_0\rangle /n_0(0)$. This 
difference is clearly very significant, and it represents the most 
important uncertainty in the extraction of $\eta_0$ from data. This 
uncertainty can only be addressed by studying more accurate solution 
of the Boltzmann equation in the case of a density dependent relaxation
time. 

%%%%%%%%%%%%%%%%%%%%%%%%%%%%%%%%%%%%%%%%%%%%%%%%%%%%%%%%%%%%%%%%%%%%%%%%
\section{Bulk viscosity}
\label{sec_bulk}
%%%%%%%%%%%%%%%%%%%%%%%%%%%%%%%%%%%%%%%%%%%%%%%%%%%%%%%%%%%%%%%%%%%%%%%%

%%%%%%%%%%%%%%%%%%%%%%%%%%%%%%%%%%%%%%%%%%%%%%%%%%%%%%%%%%%%%%%%%%%%%%%%%
\begin{figure}[t]
\bc\includegraphics[width=10cm]{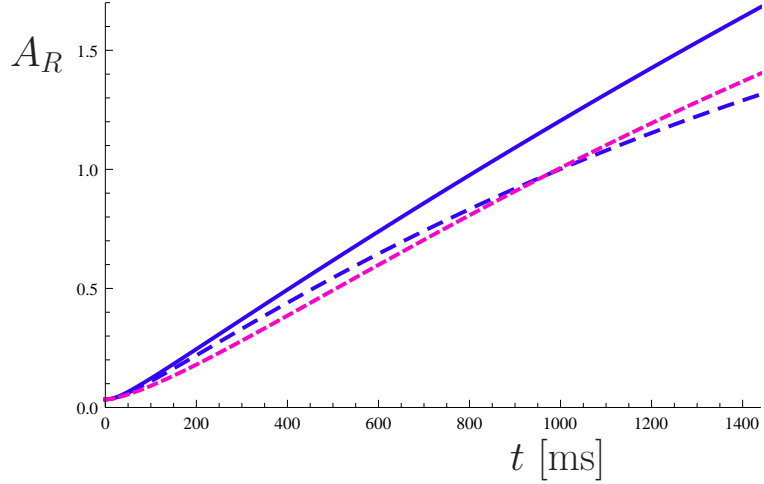}\ec
\caption{\label{fig_ar_shear_bulk}
This figure shows the effect of shear and bulk viscosity on the 
evolution of the aspect ratio $A_R(t)$ in viscous hydrodynamics.
The solid line shows the result in ideal hydrodynamics, the long
dashed line corresponds to the Navier-Stokes equation with $\beta_\eta=
0.25$, and the short dashed line shows the Navier-Stokes result for
$\beta_\zeta=0.25$. The system parameters are the same as in  
Fig.~\ref{fig_relax_vs_hydro}.}   
\end{figure}
%%%%%%%%%%%%%%%%%%%%%%%%%%%%%%%%%%%%%%%%%%%%%%%%%%%%%%%%%%%%%%%%%%%%%%%%%

%%%%%%%%%%%%%%%%%%%%%%%%%%%%%%%%%%%%%%%%%%%%%%%%%%%%%%%%%%%%%%%%%%%%%%%%%
\begin{figure}[t]
\bc\includegraphics[width=9.5cm]{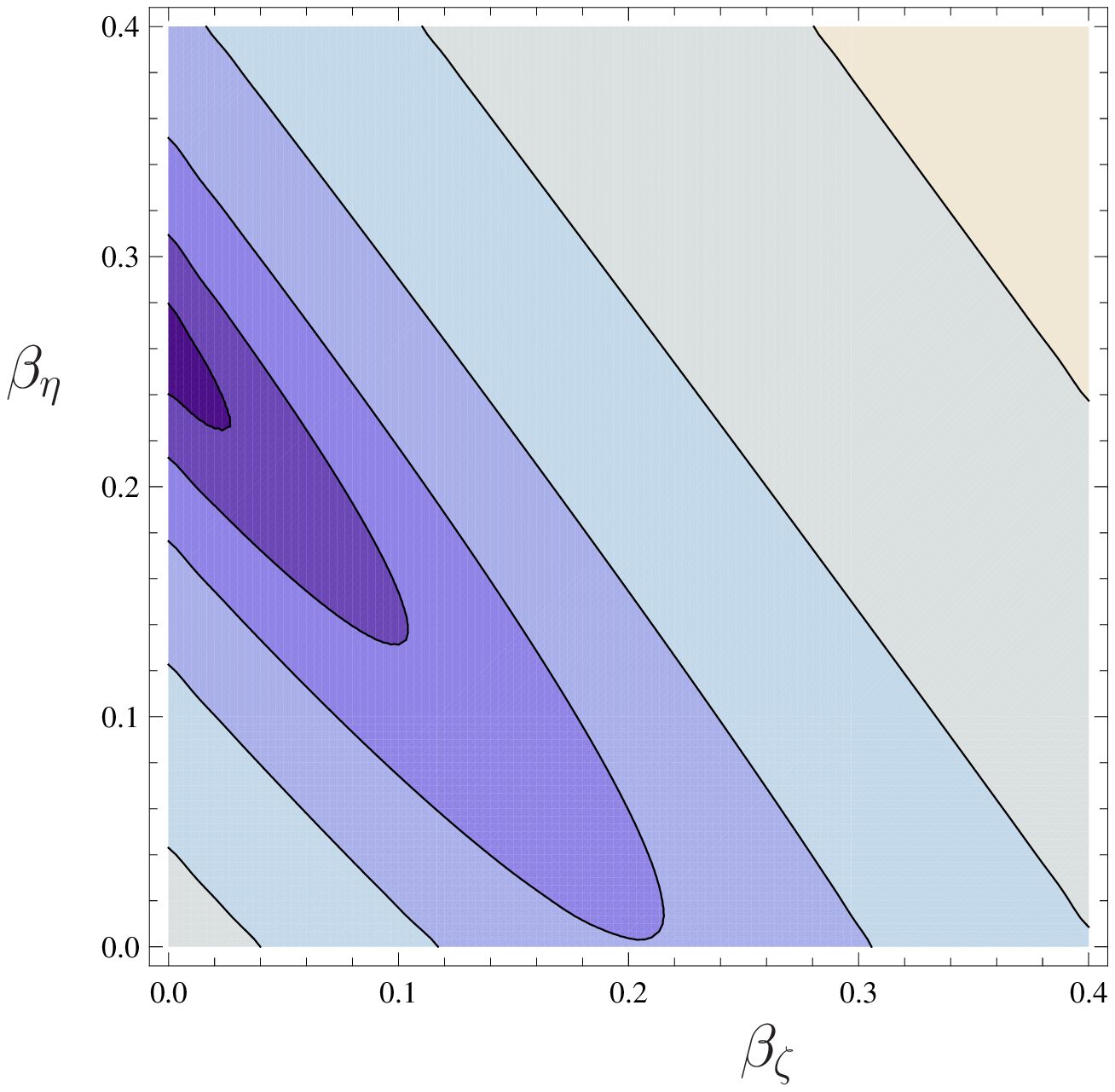}\ec
\caption{\label{fig_shear_vs_bulk}
This figure shows $\log(\chi^2)$ contours for a hydrodynamic fit 
to the data of Cao et al.~shown in Fig.~\ref{fig_relax_vs_hydro}. 
The minimum of $\chi^2$ corresponds to the dark region. The two 
parameter $\beta_\zeta$ and $\beta_\eta$ control the bulk and shear 
viscosity of the system.}   
\end{figure}
%%%%%%%%%%%%%%%%%%%%%%%%%%%%%%%%%%%%%%%%%%%%%%%%%%%%%%%%%%%%%%%%%%%%%%%%%

 Finally, we wish to study the possible effects of bulk viscosity 
on the evolution of the cloud. At unitarity we expect the bulk 
viscosity to vanish \cite{Son:2005tj}. There are nevertheless
at least two reasons for studying dissipative effects associated 
with bulk motion. The first is to verify that the theoretical 
prediction $\zeta=0$ at unitarity is indeed correct. The second 
is to understand how bulk viscosity manifests itself as one moves 
away from unitarity.

 In Sect.~\ref{sec_relax} we showed that the relaxation time 
approximation to the Boltzmann equation (without mean field 
effects) leads to a traceless dissipative stress tensor, see
equ.~(\ref{del_pi_1},\ref{del_pi_2}). We can break scale invariance 
by using a mean field potential with $\alpha\neq 2/3$. However, 
within the approximations used in Sect.~\ref{sec_mfa}, scale 
breaking in the Boltzmann-Vlasov equation does not lead to
a trace-term in the dissipative stress tensor. We will therefore 
study the role of bulk viscosity using the Navier-Stokes equation. 
We will follow \cite{Cao:2010,Schafer:2010dv} and use a scaling 
ansatz for the hydrodynamic variables, 
\be
n(x_i,t)= {\cal V}^{-1} n_0(x_i/b_i,t)\, , \hspace{1cm}
V_i= (\dot{b}_i/b_i)x_i \, , \hspace{1cm}
(\nabla_i P)/n = a_i x_i\, . 
\ee
The functional form of the density $n$ and the velocity field $V_i$ 
agree with the results obtained in Sect.~\ref{sec_relax}. In the 
Navier-Stokes limit the function $a_i(t)$ can be related to the 
parameter $\theta_i(t)$ in the distribution function by $a_i=
\theta_i\omega_i^2/b_i^2$. For an axially symmetric system the 
equations of motion for $b_i$ and $a_i$ are given by
\bea
\label{ns_for_1} 
\frac{\ddot b_\perp}{b_\perp}  &=&  a_\perp
   -  \frac{2\beta_\eta\omega_\perp}{b^2_\perp}
      \left( \frac{\dot b_\perp}{b_\perp} 
                - \frac{\dot b_z}{b_z} \right)
   -  \frac{3\beta_\zeta\omega_\perp}{b^2_\perp}
      \left(2 \frac{\dot b_\perp}{b_\perp} 
                + \frac{\dot b_z}{b_z} \right)\, ,  \\
\label{ns_for_2}
\frac{\ddot b_z}{b_z}  &=& a_z
   +  \frac{4\beta_\eta\lambda\omega_z}{b^2_z}
      \left( \frac{\dot b_\perp}{b_\perp} 
                - \frac{\dot b_z}{b_z} \right)
   -  \frac{3\beta_\zeta\lambda\omega_z}{b^2_z}
      \left( 2\frac{\dot b_\perp}{b_\perp} 
                + \frac{\dot b_z}{b_z} \right) \, ,\\
\label{ns_for_3}
\dot{a}_\perp  &=& 
 \mbox{}-\frac{2}{3}\,a_\perp
   \left(5\,\frac{\dot{b}_\perp}{b_\perp} + \frac{\dot{b}_z}{b_z}\right)
 + \frac{8\beta_\eta\omega_\perp^2}{3b^2_\perp}
  \left(\frac{\dot{b}_\perp}{b_\perp} - \frac{\dot{b}_z}{b_z}\right)^2
 + \frac{2\beta_\zeta\omega_\perp^2}{b^2_\perp}
  \left(2\frac{\dot{b}_\perp}{b_\perp} + \frac{\dot{b}_z}{b_z}\right)^2
  \, , \\
\label{ns_for_4}
 \dot{a}_z  &=& 
 \mbox{}-\frac{2}{3}\,a_z
   \left(4\,\frac{\dot{b}_z}{b_z} + 2\, \frac{\dot{b}_\perp}{b_\perp}\right)
 + \frac{8\beta_\eta\lambda\omega_z}{3b_z^2}
  \left(\frac{\dot{b}_\perp}{b_\perp} - \frac{\dot{b}_z}{b_z}\right)^2
 + \frac{2\beta_\zeta\lambda\omega_z}{b_z^2}
  \left(2\frac{\dot{b}_\perp}{b_\perp} + \frac{\dot{b}_z}{b_z}\right)^2
  \, .
\eea
Here, $\beta_\eta$ is defined in equ.~(\ref{beta_eta}) and $\beta_\zeta$
is the analogous parameter that controls the effect of bulk viscosity
\be 
\label{beta_zeta}
 \beta_\zeta = \frac{\zeta_n}{(3N\lambda)^{1/3}}\frac{1}{E_0/E_F}\, 
\ee
with $\zeta=\zeta_n n$. For $\beta_\zeta=0$ and to first order in 
$\beta_\eta$ eqns.~(\ref{ns_for_1}-\ref{ns_for_4}) are equivalent
to eqns.~(\ref{BE_b_j},\ref{BE_th_j}). Solutions of these equations
for $(\beta_\eta,\beta_\zeta)=(0.25,0)$ and $(\beta_\eta,\beta_\zeta)=
(0,0.25)$ are shown in Fig.~\ref{fig_ar_shear_bulk}. We see that 
the effects of shear and bulk viscosity are clearly distinguishable. 
Shear viscosity leads to a characteristic curvature in the time
dependence of the aspect ratio. This is related to the fact that
shear viscosity is more efficient in moving kinetic energy from 
transverse motion to longitudinal motion, combined with the fact 
that the longitudinal expansion takes place on a longer time 
scale. A comparison with the data of Cao et al.~is shown in 
Fig.~\ref{fig_shear_vs_bulk}. The figure shows $\chi^2$ contours
for a fit to the data at $E/E_f=3.61$ using both $\beta_\eta$ and 
$\beta_\zeta$ as fit parameters. We observe that the absolute
minimum is at $\beta_\eta=0.25$ and $\beta_\zeta=0.0$ ($\alpha_n
=22.1$ and $\zeta_n=0.0$). We conclude that the data at unitarity
favor a vanishing bulk viscosity and a non-zero shear viscosity.

%%%%%%%%%%%%%%%%%%%%%%%%%%%%%%%%%%%%%%%%%%%%%%%%%%%%%%%%%%%%%%%%%%%%%%%%
\section{Summary and conclusions}
\label{sec_sum}
%%%%%%%%%%%%%%%%%%%%%%%%%%%%%%%%%%%%%%%%%%%%%%%%%%%%%%%%%%%%%%%%%%%%%%%%

 In this paper we have studied a number of issues related to the 
crossover between kinetic theory and hydrodynamics. In this 
section we briefly summarize the main findings and point to 
open problems. 

\begin{enumerate}
\item Dimensional analysis implies that the shear viscosity and 
the relaxation time can be written as $\eta=\alpha_n n$ and $\tau=
\tau_n/T$, respectively. Scale invariance dictates that at 
unitarity $\alpha_n$ and $\tau_n$ are only functions of $y=
mT/n^{2/3}$. The matching between kinetic theory in the 
relaxation time approximation and hydrodynamics is simplest
in the case that $\alpha_n$ and $\tau_n$ are constant. In this 
case the matching condition is simply $\alpha_n=\tau_n$, and we 
showed that for systems that have been studied experimentally 
terms of higher order in $\tau$ are not large.

\item The Boltzmann equation describes the evolution of a system 
that obeys an ideal gas equation of state, $P=nT$. Corrections
due to a more complicated equation of state can be taken into
account using the Boltzmann-Vlasov equation and a mean field
potential. At unitarity and in the limit $\tau\to 0$ the 
mean field potential has no effect on the evolution of the 
system. As a consequence, mean field corrections remain small
even if the relaxation time is not zero. Using the mean energy 
per particle from a realistic equation of state we find that 
the uncertainty in the shear viscosity related to mean field 
effects is $\sim 2\%$. This is consistent with the effect due 
to the deviation of the equation of state from $P=nT$ found
in calculations using the Navier-Stokes equation \cite{Schafer:2010dv}.

\item We also studied the effect of a possible freezeout, that means
a transition from hydrodynamic to ballistic behavior, on the 
evolution of the system. Freezeout could occur because the 
scattering length is not strictly infinite, or because non-equilibrium
effects not captured by the Boltzmann equation cause the growth 
of the transport cross section to lag behind the equilibrium 
prediction. We showed that the experimental data obtained by 
Cao et al.~\cite{Cao:2010} require that freezeout has to 
occur late, for values of the scale parameters $b_\perp\gsim 5$. 
This implies that the unitary Fermi gas remains hydrodynamic
despite the fact that the density drops significantly during
the expansion. 

\item Finally, we studied the possible role of bulk viscosity on 
the evolution of the system. We showed that there are characteristic
differences between dissipative effects due to bulk and shear 
viscosity, and that the data are best described by shear viscosity
only.  
\end{enumerate}

 There are a number of issues that remain to be resolved. We have 
studied the Boltzmann equation with a relaxation time that scales 
as $\tau\sim T^{1/2}/n$. This is the scaling which is predicted by
the linearized form of the full collision terms using the scattering
amplitude in the unitary limit. It corresponds to a shear viscosity
that is only a function of temperature and not of density. For simplicity 
we have employed the scaling ansatz equ.~(\ref{f_scal_ans}) for the 
distribution function and solved for the coefficients by taking moments. 
We obtain an equation of motion which is equivalent to the case $\tau=
\tau_n/T$ with an effective $\tau_n=\eta(T_0)/\langle n_0\rangle $. 
This result is in qualitative, but not in quantitative, agreement 
with an estimate based on hydrodynamics with a finite relaxation
time \cite{Schaefer:2009px,Cao:2010}. In order to resolve this
disagreement we need to find more accurate solutions of the Boltzmann
equation. For infinite systems there are solutions to the linearized 
Boltzmann equation that can be matched to second order hydrodynamics
\cite{Chao:2010tk}, but the corresponding solutions for an expanding
system have not been determined. 

 We also observed that within the approximations used in this 
work the Boltzmann equation does not account for bulk viscosity 
even if scale invariance is broken by mean field corrections to 
the equation of state. It is known that the integral of the 
frequency dependent bulk viscosity is non-zero if the equation
of state violates scale invariance \cite{Taylor:2010ju}. It
is not know what the leading $1/a$ term in the bulk viscosity 
is, and what terms in the Boltzmann equation have to be kept 
in order to reproduce this term. 

 Acknowledgments: This work was supported in parts by the US 
Department of Energy grant DE-FG02-03ER41260. We thank John 
Thomas for useful discussions. 

%%%%%%%%%%%%%%%%%%%%%%%%%%%%%%%%%%%%%%%%%%%%%%%%%%%%%%%%%%%%%%%%%%%%%%%%%


\begin{thebibliography}{20}

\bibitem{Bloch:2007}
I.~Bloch, J.~Dalibard, W~Zwerger,
%``Many-Body Physics with Ultracold Gases''
Rev.\ Mod.\ Phys.\ {\bf 80}, 885 (2008)
[arXiv:0704.2511].
%%CITATION = RMPHA,80,885;%%

\bibitem{Giorgini:2008}
S.~Giorgini, L.~P.~Pitaevskii, S.~Stringari, 
%``Theory of ultracold atomic Fermi gases''
Rev.\ Mod.\ Phys.\ {\bf 80} 1215 (2008)
[arXiv:0706.3360].
%%CITATION = RMPHA,80,1215;%%

\bibitem{Chin:2009}
C.~Chin, R.~Grimm, P.~Julienne, E.~Tiesinga 
% Feshbach resonances in ultracold gases
Rev.\ Mod.\ Phys.\ {\bf 82} 1225 (2010)
[arXiv:0812.1496].
%%CITATION = RMPHA,82,1225;%%

\bibitem{Schafer:2009dj}
T.~Sch\"afer and D.~Teaney,
%``Nearly Perfect Fluidity: From Cold Atomic Gases to Hot Quark Gluon
%Plasmas,''
Rept.\ Prog.\ Phys.\  {\bf 72}, 126001 (2009)
[arXiv:0904.3107 [hep-ph]].
%%CITATION = ARXIV:0904.3107;%%

\bibitem{OHara:2002}
%``Observation of a Strongly-Interacting Degenerate Fermi Gas of Atoms''
K.~M.~O'Hara, S.~L.~Hemmer, M.~E.~Gehm, S.~R.~Granade, J.~E.~Thomas,
Science Vol.\ 298, No.\ {\bf 5601}, 2179 (2002)
[cond-mat/0212463].
%%CITATION = COND-MAT 0212463;%%

\bibitem{Schafer:2007pr}
T.~Sch\"afer,
%``The Shear Viscosity to Entropy Density Ratio of Trapped Fermions in the
%Unitarity Limit,''
Phys.\ Rev.\  A {\bf 76}, 063618 (2007)
[arXiv:cond-mat/0701251].
%%CITATION = PHRVA,A76,063618;%%

\bibitem{Turlapov:2007}
%``Is a Gas of Strongly Interacting Atomic Fermions a Nearly Perfect Fluid''
A.~Turlapov, J.~Kinast, B.~Clancy, L.~Luo, J.~Joseph, J.~E.~Thomas,
J.\ Low Temp.\ Phys.\ {\bf 150}, 567 (2008)
[arXiv:0707.2574].
%%CITATION = ARXIV:0707.2574;%%

\bibitem{Schaefer:2009px}
T.~Sch\"afer and C.~Chafin,
%``Scaling Flows and Dissipation in the Dilute Fermi Gas at Unitarity,''
arXiv:0912.4236 [cond-mat.quant-gas].
%%CITATION = ARXIV:0912.4236;%%

\bibitem{Thomas:2009zz}
J.~E.~Thomas,
%``Is An Ultra-Cold Strongly Interacting Fermi Gas A Perfect Fluid?,''
Nucl.\ Phys.\  A {\bf 830}, 665C (2009). 
%%CITATION = NUPHA,A830,665C;%%

\bibitem{Cao:2010}
C.~Cao, E.~Elliott, J.~Joseph, H.~Wu, J.~Petricka, 
T. Sch{\"a}fer, J.~E.~Thomas,
%``Observation of Universal Temperature Scaling in the Quantum 
% Viscosity of a Unitary Fermi Gas''
arXiv:1007.2625 [cond-mat.quant-gas].
%%CITATION = ARXIV:1007.2625;%%

\bibitem{Kovtun:2004de}
P.~Kovtun, D.~T.~Son and A.~O.~Starinets,
%``Viscosity in strongly interacting quantum field theories from black hole
%physics,''
Phys.\ Rev.\ Lett.\  {\bf 94}, 111601 (2005)
[arXiv:hep-th/0405231].
%%CITATION = PRLTA,94,111601;%%

\bibitem{rhic:2005}
I.~Arsenne et al.~[Brahms],
B.~Back et al.~[Phobos],
K.~Adcox et al.~[Phenix],
J.~Adams et al.~[Star],
``First Three Years of Operation of RHIC'',
Nucl.\ Phys.\ {\bf A757}, 1-183 (2005).

\bibitem{Romatschke:2007mq}
P.~Romatschke and U.~Romatschke,
%``Viscosity Information from Relativistic Nuclear Collisions: How Perfect is
%the Fluid Observed at RHIC?,''
Phys.\ Rev.\ Lett.\  {\bf 99}, 172301 (2007)
[arXiv:0706.1522 [nucl-th]].
%%CITATION = PRLTA,99,172301;%%

\bibitem{Dusling:2007gi}
K.~Dusling and D.~Teaney,
%``Simulating elliptic flow with viscous hydrodynamics,''
Phys.\ Rev.\  C {\bf 77}, 034905 (2008)
[arXiv:0710.5932 [nucl-th]].
%%CITATION = PHRVA,C77,034905;%%

\bibitem{Song:2008hj}
H.~Song and U.~W.~Heinz,
%``Extracting the QGP viscosity from RHIC data -- a status report from viscous
%hydrodynamics,''
arXiv:0812.4274 [nucl-th].
%%CITATION = ARXIV:0812.4274;%%

\bibitem{Guery:1999}
D.~Guery-Odelin, F.~Zambelli, J.~Dalibard, and S.~Stringari,  
%``Collective oscillations of a classical gas confined in harmonic traps''
Phys.\ Rev.\ A {\bf 60} 4851 (1999). 

\bibitem{Pedri:2002}
P.~Pedri, D.~Guery-Odelin and S.~Stringari,  
%``Dynamics of a classical gas including dissipative and mean-field effects''
Phys.\ Rev.\ A {\bf 68} 043608 (2003)
[ond-mat/0305624]

\bibitem{Guery:2002}
D.~Guery-Odelin,
%``Mean-field effects in a trapped gas''
Phys.\ Rev.\ A {\bf 66} 033613 (2002).

\bibitem{Menotti:2002}
%``Expansion of an interacting Fermi gas''
C.~Menotti, P.~Pedri, S.~Stringari,
Phys.\ Rev.\ Lett.\ {\bf 89}, 250402 (2002)
[cond-mat/0208150].
%%CITATION = PRLTA,89,250402;%%

\bibitem{Jackson:2004}
B.~Jackson, P.~Pedri, and S.~Stringari,
%``Collisions and expansion of an ultracold dilute Fermi gas''
Europhys.\ Lett.\ {\bf 67} 524 (2004).

\bibitem{Nikuni:2004}
T.~Nikuni, A.~Griffin
%``Frequency and damping of hydrodynamic modes in a Bose-condensed gas''
Phys.\ Rev.\ A {\bf 69}, 023604 (2004)
[arXiv:cond-mat/0309269].

\bibitem{Bruun:2005}
% ``Viscosity and thermal relaxation for a resonantly interacting 
% Fermi gas''
G.~M.~Bruun, H.~Smith,
Phys.\ Rev.\ A {\bf 72}, 043605 (2005) 
[cond-mat/0504734].
%%CITATION = COND-MAT 0504734;%%

\bibitem{Bruun:2006}
% ``Shear viscosity and damping for a Fermi gas in the unitarity limit''
G.~M.~Bruun, H.~Smith,
Phys.\ Rev.\ A {\bf 75}, 043612 (2007)
[cond-mat/0612460].
%%CITATION = COND-MAT 0612460;%%

\bibitem{Bruun:2007}
%``Frequency and damping of the Scissors Mode of a Fermi gas''
G.~M.~Bruun, H.~Smith
Phys. Rev. A {\bf 76}, 045602 (2007) 
[arXiv:0709.1617].

\bibitem{Son:2005tj}
D.~T.~Son,
%``Vanishing bulk viscosities and conformal invariance of unitary Fermi gas,''
Phys.\ Rev.\ Lett.\  {\bf 98}, 020604 (2007)
[arXiv:cond-mat/0511721].
%%CITATION = PRLTA,98,020604;%%

\bibitem{Enss:2010qh}
T.~Enss, R.~Haussmann, W.~Zwerger,
%``Viscosity and scale invariance in the unitary Fermi gas,''
Annals Phys.\  {\bf 326}, 770-796 (2011).
[arXiv:1008.0007 [cond-mat.quant-gas]].

\bibitem{Schafer:2010dv}
T.~Sch\"afer,
%``Dissipative fluid dynamics for the dilute Fermi gas at unitarity: 
% Free expansion and rotation,''
Phys.\  Rev.\ A (2010), in press
[arXiv:1008.3876 [cond-mat.quant-gas]].

\bibitem{Thomas:2005}
J.~E.~Thomas, J.~Kinast, and A.~Turlapov, 
%``Virial theorem and universality of a unitary Fermi gas''
Phys.\ Rev.\ Lett.\ {\bf 95}, 120402 (2005) [arXiv:cond-mat/0503620].

\bibitem{Werner:2006}
F.~Werner and Y.~Castin, 
%``Unitary gas in an isotropic harmonic trap: symmetry properties
% and applications'',
Phys.\ Rev.\ A {\bf 74}, 053604 (2006) [cond-mat/0607821].

\bibitem{Son:2007}
D.~T.~Son, 
%``Three comments on the Fermi gas at unitarity in a harmonic trap''
arXiv:0707.1851.[cond-mat.other].

\bibitem{Chevy:2009}
S.~Nascimbene, N.~Navon, K.~Jiang, F.~Chevy, C~Salomon,
%``Exploring the Thermodynamics of a Universal Fermi Gas''
Nature {\bf 463}, 1057 (2010)
[arXiv:0911.0747[cond-mat.quant-gas]].
%%CITATION = ARXIV:0911.0747;%%

\bibitem{Chao:2010tk}
J.~Chao, M.~Braby, T.~Sch\"afer,
%``Viscosity spectral functions of the dilute Fermi gas in kinetic theory,'' 
[arXiv:1012.0219 [cond-mat.quant-gas]].

\bibitem{Taylor:2010ju}
E.~Taylor and M.~Randeria,
%``Viscosity of strongly interacting quantum fluids: spectral functions and
%sum rules,''
Phys.\ Rev.\  {\bf A81}, 053610 (2010).
[arXiv:1002.0869 [cond-mat.quant-gas]].
%%CITATION = ARXIV:1002.0869;%%

\end{thebibliography}
\end{document}